\documentclass[11pt,thmsa]{article}
\usepackage{amssymb}

\usepackage{sw20lart}

\input{tcilatex}
\begin{document}

\smallskip 
\begin{titlepage}
\begin{center}
\vspace{1cm}
\hfill
\vbox{
    \halign{#\hfil         \cr
         hep-th/0110005 \cr
           IPM/P-2001/033\cr
           Sept. 2001\cr} }
      
\vskip 1cm
{\Large \bf
Noncommutative Supersymmetry in Two Dimensions  }
\vskip 0.5cm
{\bf Reza Abbaspur\footnote{e-mail:abbaspur@theory.ipm.ac.ir} 
}\\ 
\vskip .25in
{\em
Institute for Studies in Theoretical Physics and Mathematics,  \\
P.O. Box 19395-5531,  Tehran,  Iran.\\}
\end{center}
\vskip 0.5cm

\begin{abstract}

Based on an argument for the noncommutativity of momenta in noncommutative
directions, we arrive at a generalization of the ${\cal N}=1$ super $E^2$ algebra associated
to the deformation of translations in a noncommutative Euclidean plane. The algebra is obtained 
using appropriate representaions of its generators on the space of superfields in a 
$D=2,  {\cal N}=1$ ``noncommutative superspace.'' We find that the (anti)commutators 
between several (super)translation generators are no longer vanishing, but involve a new 
set of generators which together with the (super)translation and rotation generators form a 
consistent closed algebra. We then analyze the spectrum of this algebra in order to obtain 
its fundamental and adjoint representations.

\end{abstract}

\end{titlepage}\newpage 

\section{\protect\smallskip Introduction}

Supersymmetry (SUSY) has been the basic ingredient in most of the recent
attempts towards formulating a unified fundamental theory of the nature.
Besides its merit for solving some important theoretical shortcomings of the
standard particle physics, it has also played a central role in recent
developments of string theory \cite{1}. For being a consistent fundamental
description of nature, string theory depends on SUSY at least at the level
of the string worldsheet. This in turn leads to a few number of consistent
spacetime SUSY theories including all the well known types of superstring
theories. All these string theories at their low energy level reduce to SUSY
effective field theories from which the supergravity and super Yang-Mills
(SYM) theories are the most famous examples \cite{2}.

\smallskip

One of the important consequences of string theory is the emergence of the
spacetime noncommutativity at short distances (for a recent review see \cite
{3.5}). \smallskip It is shown that this noncommutativity of spacetime is
equivalent to deforming the algebra of functions on the spacetime manifold
by replacing their ordinary products with the so called Moyal products. This
type of noncommutativity was initially appeared in the context of Matrix
theory compactifications on a torus \cite{3.1, 3.2}. The more recent
occasion of noncommutativity in string theory was that for the open strings
ending on a D-brane within a background B-field \cite{3.3}. \smallskip In
both cases, the low energy limit turns out to be a noncommutative gauge
theory which is expected to be supersymmetric because of the SUSY underlying
the fundamental theory. Thus we are led to NCSYM theory as the worldvolume
theory of a D-brane.

\smallskip

It is well known that BPS D-branes of string theory are SUSY extended
objects preserving 1/2 of the total spacetime SUSY and they possess the same
amount of supercharges on their worldvolume theories \cite{2}. One may
suspect that whether the amount of SUSY on a brane is changed by the
presence of a B-field, and further, if a B-field modifies the algebra of the
supercharges on the brane. A direct computation of the worldsheet
(super)charges reveals that the answer to both questions is negative; i.e.
the presence a (constant parallel) B-field does not alter the SUSY
properties of a D-brane \cite{4.1} . This leaves open, however, the
possibility of a deformation of the SUSY algebra in a more general given
background. In fact, in \cite{4.2} it is shown that a constant H-field ($%
H=dB $) deforms the spacetime superalgebra, using explicit expressions for
the conserved worldsheet (super)charges.

\smallskip

\smallskip

Whether the prediction of string theory is about SUSY in the presence of
background fields, it is worthwhile in its own respect to know how the
spacetime noncommutativity may influence the superalgebra underlying a SUSY
field theory. The first attempts in this direction were made in refs.\cite
{4.1, 5.1, 6.1} (see also \cite{5.2}-\cite{5.4} for more recent works). The
deformation proposed in \cite{5.1} was based on the deformation of the
algebra of superfields on a so-called ``noncommutative superspace''. This is
defined by the algebra of the bosonic and fermionic operators $(x^{\mu
},\theta ^{\alpha })$ (corresponding to the grassmann even and odd)
coordinates of the ordinary superspace obeying non-trivial (anti)commutation
relations among themselves. This leads to a new generalized version of the
Moyal product between superfields involving a deformation of the products of
functions of the odd coordinates similar to that of the even coordinates
(for a general deformed version of the grassmann algebra\smallskip see \cite
{5.5}, and \cite{5.6} for a non-anticommutative field theory). It was found,
however, that (at least for the $D=4,$ $\mathcal{N}=1$ case) the only
deformation compatible with supertranslation and closure of the chiral
superfields under the generalized Moyal product is the one in which $\theta $%
's anticommute with themselves and commute with $x$'s, as usual, but $x$'s
can still be noncommuting.

\smallskip

On the basis of the above result the authors of \cite{5.1, 6.1} have
constructed deformed versions of the usual Wess-Zumino model and SYM\ theory
with $\mathcal{N}=1$ SUSY in four dimensions. Deformations of the $\mathcal{N%
}=2,4$ SYM theories have been studied in refs. \cite{6.11, 6.12, 6.13} and
that of SUSY Born-Infeld action in \cite{5.1, 6.10, 6.9}, and the deformed
version of the non-linear sigma-models was considered in \cite{6.8, 6.7}. In
all these cases, the deformed theory is obtained from its commutative
counterpart simply by replacing the ordinary products of fields with the
Moyal products. The renormalizibility and 1-loop effective actions of SUSY
NCFT was studied in refs. \cite{6.2}-\cite{6.13}. Typically,
noncommutativity turns a renormalizable theory to a non-renormalizable
theory due to the UV/IR mixing phenomenon \cite{6.15}. However, the addition
of SUSY restores the renormalizibility of the theory and moreover results in
the non-renormalization of its parameters \cite{6.2}-\cite{6.13}.

\smallskip

\smallskip

\smallskip More recently, some authors have considered the issue of
noncommutativity in quantum mechanics (QM) \cite{7.1}-\cite{7.8}. The main
problem of the NC-QM is to find the spectrum of states for a particle in a
specific potential assuming that its coordinates $(q^{i})$ do not commute.
The momenta of the particle $(p_{i})$ may be commuting or non-commuting
depending on the given problem. Generically, the commutation relations of $%
(q^{i},p_{i})$ have the following form: 
\begin{equation}
\lbrack q^{i},q^{j}]=i\Theta ^{ij},\quad [q^{i},p_{j}]=i\delta
_{j}^{i},\quad [p_{i},p_{j}]=i\Phi _{ij}.\smallskip  \label{1}
\end{equation}
This situation is somehow familiar from the ordinary QM of charged particles
on a plane in a constant (transverse) magnetic field and is well known as
the Landau problem \cite{7.1}. Of course, for the ordinary (commutative)
Landau problem we have $\Theta ^{ij}=0,\;\Phi _{ij}\neq 0$. It is easy to
show, by a suitable linear transformation on $(q^{i},p_{i}),$ that this
problem can be mapped to a problem in the noncommutative space, i.e. the one
with $\Theta ^{ij}\neq 0,\;\Phi _{ij}=0$. Such a transformation is indeed
the essence of the Seiberg-Witten map relating a commutative gauge theory
with a B-field to a noncommutative gauge theory without a B-field\footnote{%
As discussed in \cite{3.4}, the parameters $\Phi _{ij}$ in eqs.(\ref{1}) are
the same that appear in the formulation of a NC-DBI action as a freedom to
interpolate between several equivalent descriptions of the same theory \cite
{3.3}.} \cite{3.3}. In general, for arbitrary $\Theta ,\Phi $, one can find
linear transformations on $(q^{i},p_{i})$ in a way transforming their
commutation relations to the canonical form; i.e. the one with $\Theta
^{ij}=\Phi _{ij}=0$. In such a situation, we deal essentially with an
ordinary QM problem. There is, however, a critical value of the B-field, $%
\Phi =-\Theta ^{-1}$, for which the canonizing linear transformation on $%
(q^{i},p_{i})$ becomes singular. In that case the phase space of the
particle becomes degenerate and is actually described by half of the
coordinates $(q^{i},p_{i})$ (say by $q^{i}$ only) and the density of states
on the phase space becomes infinity \cite{7.3}. This means that in the
critical case, the momenta of a particle are no longer independent of its
coordinates but related to them in a particular way\cite{3.4, 7.3}. This is
the same situation occurring for a particle in a large magnetic field or in
its first Landau level \cite{7.1, 7.2}, which is usually treated using the
Dirac quantization procedure for the constrained systems \cite{7.1}.

\smallskip

\smallskip

In the critical case $\Phi =-\Theta ^{-1}$, both $q^{i}$ and $p_{i}$ are
noncommuting and we can not get rid of the noncommutativity by a linear
transformation on $(q^{i},p_{i})$. In the context of the noncommutative
gauge theory, this is equivalent to the fact that, for the critical value of
the noncommutative field strength, $\widehat{F}=\Theta ^{-1}$, one can not
find a commutative description of the same theory \cite{3.3}. This is the
same choice which is naturally singled out by the matrix theory \cite{3.4}.
As explained in \cite{3.4}, in this case the derivative operators $\partial
_{i}$ on functions of the noncommutative space are expressed as commutators
with the coordinates and hence they obey the noncommutativity relation $%
[\partial _{i},\partial _{j}]=-i\Phi _{ij}=i\Theta _{ij}^{-1}$. This means
that successive translations in a noncommutative space generally do not
commute. When extended to a noncommutative superspace, this raises the
question that how the SUSY algebra is affected by turning on the spacetime
noncommutativity. It is the purpose of this paper to answer to this question.

\smallskip

In this paper we are mainly interested in the critical ($\Phi =-\Theta ^{-1}$%
) regime in two (Euclidean) dimensions, although most of the results seem to
have natural generalizations to the non-critical case and in higher
dimensions. More specifically, we are interested in the noncommutative
deformation of the $\mathcal{N}=1$ superalgebra corresponding to a
deformation of the two dimensional Euclidean group (briefly denoted as $%
E_{\vartheta }^{2}$). The cases of higher dimensional spaces and of the
extended SUSY will be investigated elsewhere \cite{8}.

\smallskip

The paper is organized as follows. In section 2 we will briefly review the
basic notions of superspace and supersymmetry in the commutative theories
and then introduce the spacetime noncommutativity and the operator
formalism. Then in section 3 we derive the deformed super $E^{2}$ algebra
corresponding to a noncommutative Euclidean 2D space. The problem of
representation and spectrum of the deformed superalgebra is addressed in
section 4 and we conclude the paper by a summary and discussion in section 5.

\smallskip

\section{Preliminaries and Notations}

In this section we first briefly review some elements of SUSY in commutative
spaces and then introduce the spacetime noncommutativity and its associated
operator formalism. This will fix our notation for the rest of this paper.

\subsection{Superspace, Superfields and Ordinary Supersymmetry}

Superspace is a mathematical tool invented for simplifying formulation of
the SUSY field theories and provides a framework to treat an internal
symmetry of a theory in parallel to its spacetime symmetries. The ordinary
superspace is defined by the algebra generated by the bosonic and fermionic
coordinates $(x^{a},\theta ^{\alpha })$ satisfying the following
(anti)commutation relations 
\begin{equation}
\lbrack x^{a},x^{b}]=0,\quad \{\theta ^{\alpha },\theta ^{\beta }\}=0,\quad
[x^{a},\theta ^{\alpha }]=0.  \label{2}
\end{equation}
A general superfield is then an arbitrary function of $(x^{a},\theta
^{\alpha })$. For $\mathcal{N}$ $=1,\,D=2$ superspace, we have 2 bosonic and
2 fermionic coordinates and so the indices range as $a=1,2$ and $\alpha
=+,-. $ (see also Appendix for conventions on spinors). A general superfield
on this superspace has the following expansion 
\begin{eqnarray}
\mathcal{S}(x,\theta ) &=&\phi (x)+\overline{\theta }\psi (x)+\frac{1}{2}%
\overline{\theta }\theta F(x)  \nonumber \\
&=&\phi (x)+\theta ^{+}\psi _{+}(x)+\theta ^{-}\psi _{-}(x)+\theta
^{+}\theta ^{-}F(x).  \label{3}
\end{eqnarray}
For a scalar superfield, $\phi (x),F(x)$ are bosonic scalar fields while $%
\psi _{\pm }(x)$ form components of a fermionic spinor field. \smallskip
Obviously, the set of all superfields is closed under the ordinary product
of superfields with the above (anti)commutation rules.

For later purposes, it is useful to remind the dimensionalities of $\mathcal{%
S},\,\,\theta ,\,x,$ which in terms of the length units are written as 
\begin{equation}
\lbrack \mathcal{S}]=0,\qquad [\theta ]=\frac{1}{2},\qquad [x]=1.  \label{4}
\end{equation}

\smallskip \smallskip \smallskip The ordinary SUSY can be easily formulated
using the concept of ``translations'' in a superspace. There are two types
of these translations which are uniquely determined using dimensional
analysis as follows:

\smallskip

1. \textit{Ordinary translation}:

This is a constant shift of $x$ with the vector parameter $a$ 
\begin{equation}
\delta _{a}x^{a}=a^{a},\quad \delta _{a}\theta ^{\alpha }=0.  \label{5}
\end{equation}

\smallskip

2.\textit{\ Supertranslation}:

This is a constant shift of $\theta $ with the spinor parameter $\epsilon $%
\begin{equation}
\delta _{\epsilon }\theta ^{\alpha }=\epsilon ^{\alpha },\quad \delta
_{\epsilon }x^{a}=\overline{\epsilon }\rho ^{a}\theta .  \label{6}
\end{equation}
The appropriate superspace representations of the generators $%
P_{a},Q_{\alpha }$ of these transformations are simply determined by
examining their effects on a general superfield as follows 
\begin{eqnarray}
\delta _{a}\mathcal{S} &=&a^{a}\frac{\partial \mathcal{S}}{\partial x^{a}}%
\equiv i(a^{a}P_{a})\mathcal{S},  \nonumber \\
\delta _{\epsilon }\mathcal{S} &=&\epsilon ^{\alpha }\frac{\partial \mathcal{%
S}}{\partial \theta ^{\alpha }}+\overline{\epsilon }\rho ^{a}\theta \frac{%
\partial \mathcal{S}}{\partial x^{a}}\equiv (\overline{\epsilon }Q)\mathcal{S%
}.  \label{7}
\end{eqnarray}
These imply that 
\begin{eqnarray}
P_{a} &=&-i\frac{\partial }{\partial x^{a}},  \nonumber \\
Q_{\alpha } &=&\frac{\partial }{\partial \theta ^{\alpha }}+i(\rho ^{1}\rho
^{a})_{\alpha \beta }\theta ^{\beta }\frac{\partial }{\partial x^{a}}.
\label{8}
\end{eqnarray}
Now, using the following properties of the commutative superspace 
\begin{equation}
\lbrack \frac{\partial }{\partial x^{a}},\frac{\partial }{\partial x^{b}}%
]=0,\quad [\frac{\partial }{\partial x^{a}},\frac{\partial }{\partial \theta
^{\alpha }}]=0,\qquad \{\frac{\partial }{\partial \theta ^{\alpha }},\frac{%
\partial }{\partial \theta ^{\beta }}\}=0,  \label{9}
\end{equation}
together with the above expressions of $P,Q$, we obtain the ordinary
supertranslation algebra as follows: 
\begin{eqnarray}
\lbrack P_{a},P_{b}] &=&0,\quad [P_{a},Q_{\alpha }]=0,  \nonumber \\
\{Q_{\alpha },Q_{\beta }\} &=&-2(\rho ^{1}\rho ^{a})_{\alpha \beta }P_{a}.
\label{10}
\end{eqnarray}
\smallskip This indeed defines a subalgebra of the complete super $E^{2}$
algebra. The latter is obtained by including in the above algebra the
commutators of the $SO(2)$ generator with these (super)translation
generators.

\smallskip

The important property of a SUSY transformation is that it changes any
superfield to a similar superfield. This means that we can re-arrange $%
\delta _{\epsilon }\mathcal{S}$ of eq.(\ref{7}) in powers of $\theta ^{\pm }$
to obtain 
\begin{equation}
\delta _{\epsilon }\mathcal{S}=\delta _{\epsilon }\phi (x)+\theta ^{+}\delta
_{\epsilon }\psi _{+}(x)+\theta ^{-}\delta _{\epsilon }\psi _{-}(x)+\theta
^{+}\theta ^{-}\delta _{\epsilon }F(x),
\end{equation}
from which it follows 
\begin{eqnarray}
\delta _{\epsilon }\phi (x) &=&\overline{\epsilon }\psi (x),  \nonumber \\
\delta _{\epsilon }\psi (x) &=&\left( -\rho ^{a}\partial _{a}\phi
(x)+F(x)\right) \epsilon ,  \nonumber \\
\delta _{\epsilon }F(x) &=&-i\overline{\epsilon }\rho ^{a}\partial _{a}\psi
(x).
\end{eqnarray}
These are the familiar SUSY transformations of the component fields
equivalent to the second eq.(\ref{7}).

\subsection{Noncommutativity of Spacetime}

\textit{Origin: }

Noncommutative field theories (and in particular NCYM theory) arise in the
so called decoupling limit of the theory of open strings ending on a D-brane
in the presence of a background B-field \cite{3.3}. String theory
calculation of the open string tree amplitudes shows that the low energy
dynamics of D-branes in the presence of such a background is modified in a
way that all the ordinary products of fields in the usual theory are
replaced by the following Moyal (or $*$-) product \cite{3.3} 
\begin{equation}
f(x)*g(x)=\exp (\frac{i}{2}\Theta ^{\mu \nu }\partial _{\mu }\partial _{\nu
}^{\prime })f(x)g(x^{\prime })\mid _{x=x^{\prime }}.  \label{11}
\end{equation}
\smallskip Here, $\Theta ^{\mu \nu }=-\Theta ^{\nu \mu }$ is the constant
noncommutativity tensor which is determined by the closed string background
fields. The $*$-product is clearly a noncommutative operation; i.e. $f*g\neq
g*f$ for\thinspace general $f,g$. In particular, noncommutativity of the
spacetime coordinates is measured by $\Theta ^{\mu \nu }$ according to 
\begin{equation}
\lbrack x^{\mu },x^{\nu }]_{*}=i\Theta ^{\mu \nu }.  \label{12}
\end{equation}
It can be shown that, for a \textit{constant} $\Theta ^{\mu \nu },$ the
Moyal product is associative which means that for every triplet of functions 
$f(x),\,g(x),\,h(x)$ we have 
\begin{equation}
(f*g)*h=f*(g*h).  \label{13}
\end{equation}
It turns out that the constancy of $\Theta $ (corresponding to a flat closed
string background) is essential in proving the associativity of the $*$%
-product. Otherwise, it is modified to a so-called Kontsevich product which
in general defines a non-associative algebra \cite{9}. Note also that the
above associativity property is crucial to the Moyal-Weyl correspondence and
hence to the operator formalism which we use as a basic tool during this
paper (see below).

\smallskip

\textit{The Moyal-Weyl Correspondence:}

\smallskip The noncommutativity and associativity of the $*$-product suggest
that there must be an isomorphism between the algebra of functions on a
noncommutative space and that of the operators on some Hilbert space \cite
{10}. To see this isomorphism explicitly, let us associate to the
coordinates $x^{\mu }$ the operators $\widehat{x}^{\mu }$ satisfying the
algebra 
\begin{equation}
\lbrack \widehat{x}^{\mu },\widehat{x}^{\nu }]=i\Theta ^{\mu \nu }.
\label{14}
\end{equation}
Any function $f(x)$ is then associated to an operator $\widehat{O}_{f}=%
\widehat{f}(\widehat{x})$ (and vice versa) through a one-to-one map defined
as 
\begin{equation}
f(x)=\int d^{n}k\,\,e^{ik.x}\widetilde{f}(k)\longleftrightarrow \widehat{f}(%
\widehat{x})=\int d^{n}k\,\,e^{ik.\widehat{x}}\widetilde{f}(k),  \label{15}
\end{equation}
where $\widetilde{f}(k)$ denotes the Furrier transform of $f(x)$. This is
called the Moyal-Weyl correspondence. It is easy then to show \cite{10} the
following properties 
\begin{eqnarray}
O_{f*g} &=&O_{f}\cdot O_{g}\,,  \nonumber \\
\int d^{n}x\,f\,(x) &=&(2\pi )^{n/2}\sqrt{\det \Theta }\,\text{Tr}O_{f}\,.
\label{16}
\end{eqnarray}
These are the basis of the operator formulation of NCFT's which has played a
central role in the development of solitons of these theories \cite{10}. For
our purpose, the Moyal-Weyl correspondence is used as a basis for treating
functions of a noncommutative space, on equal footing with the symmetry
generators, as the operators of some Hilbert space.

\section{Noncommutative Supersymmetry}

\textit{Motivation:}

The issue of supersymmetry in NCFT has been considered previously by several
authors\cite{6.1}-\cite{6.7}. In particular, many perturbative aspects of
the noncommutative super Yang-Mills (NCSYM) theory have been studied in many
papers (for an exhaustive list of references see \cite{3.5}). In most of
these papers, however, the issue of supersymmetry in NCFT is treated in
parallel to what usually is done in their commutative counterpart theories.
In other words, one starts with a usual commutative theory based on an
ordinary SUSY algebra and then proceed to deform it by replacing the
ordinary products of fields with Moyal products. In this way we obtain a
noncommutative supersymmetric field theory whose field content obeys the
ordinary SUSY transformation properties.\smallskip

From a more fundamental point of view, however, we know that the SUSY
algebra underlying a theory is obtained by extending its usual spacetime
symmetries including the translations and rotations (or boosts) by inclusion
of the supertranslation symmetry. Therefore, one should expect a deformation
of SUSY, if the theory allows for a deformation of either of the usual
spacetime symmetries. In the case of a NCFT, we know that generally
noncommutativity breaks rotational symmetries of the theory but preserves
its translational symmetries. On the other hand, we know that there is a
freedom in the formulation of a NCFT due to the freedom in the choice of the
commutator of any two spatial derivatives; $[\partial _{a},\partial
_{b}]=-i\Phi _{ab}$ \cite{3.4}. Interpreting these derivatives as the
generators of translations in the noncommutative directions, we find that
the translation algebra is deformed by a central extension. Hence we should
expect also a deformation of the corresponding supertranslation algebra.

\smallskip

\smallskip

\smallskip

\subsection{Deformed Translation Algebra}

Let us consider a noncommutative 2D space with coordinates $\widehat{x}^{a}$
satisfying \footnote{%
In due course we will omit the hat signs on the operators whenever there is
no possibility of confusion.} 
\begin{equation}
\lbrack \widehat{x}^{a},\widehat{x}^{b}]=i\vartheta \epsilon ^{ab}.
\label{17}
\end{equation}
In the operator formulation, derivatives of a field $\widehat{\phi }(%
\widehat{x})$ can be expressed as \cite{3.4, 7.3} 
\begin{equation}
\partial _{a}\widehat{\phi }(\widehat{x})=i\vartheta ^{-1}\epsilon _{ab}[%
\widehat{x}^{b},\widehat{\phi }(\widehat{x})].  \label{18}
\end{equation}
This suggests identifying the momenta $\widehat{P}_{a}$ as the generators of
translations in terms of coordinates as follows (see also \cite{7.4} for
this point) 
\begin{equation}
\widehat{P}_{a}=\vartheta ^{-1}\epsilon _{ab}\widehat{x}^{b}.  \label{19}
\end{equation}
By the eq.(\ref{17}), this relation automatically satisfies the Heisenberg
equation 
\begin{equation}
\lbrack \widehat{x}^{a},\widehat{P}_{b}]=i\delta _{b}^{a}.  \label{20}
\end{equation}
\smallskip Also from the same constraint it follows that 
\begin{equation}
\lbrack \widehat{P}_{a},\widehat{P}_{b}]=i\vartheta ^{-1}\epsilon _{ab}\,,
\label{21}
\end{equation}
which shows the noncommutativity of momenta in the noncommuting directions.
The three eqs.(\ref{17}), (\ref{20}), (\ref{21}) define a deformation of the
ordinary Heisenberg algebra on a noncommutative plane in a way consistent
with all the Jacobi Identities. \smallskip

\smallskip

We can reverse this logic in order to conclude eq.(\ref{19}) in another way.
We may start by not assuming \smallskip any relation between ($\widehat{x},%
\widehat{P}$)'s but instead assume a commutation relation between $\widehat{P%
}$'s as follows 
\begin{equation}
\lbrack \widehat{P}_{a},\widehat{P}_{b}]=i\omega \epsilon _{ab}.  \label{22}
\end{equation}
where $\omega $ is some real constant. This corresponds to a more general
deformation of the Heisenberg algebra involving two arbitrary parameters%
\textit{\ }($\vartheta ,\omega $). The special point $\omega =1/\vartheta $
corresponds to the previous deformation. For generic values of ($\vartheta
,\omega $), the representation space of the deformed Heisenberg algebra is
spanned by a 2-parameter set of states with parameters defined by the
eigenvalues of the operators in a Cartan subalgebra of ($\widehat{x},%
\widehat{P}$)'s. This can be chosen to be ($\widehat{x}^{1},\widehat{P}_{2}$%
) and hence the Hilbert space is spanned by the simultaneous eigenstates $%
|x^{1},p_{2}\rangle $ of these two operators. In this basis, $(\widehat{x}%
^{1},\widehat{P}_{2})$ have the following simple representations 
\begin{equation}
\widehat{x}^{1}=x^{1},\qquad \,\,\,\widehat{P}_{2}=p_{2}.  \label{23}
\end{equation}
Comparison to the case of an ordinary Heisenberg algebra then suggests that
the other two operators $(\widehat{x}^{2},\widehat{P}_{1})$ should have
linear representations in terms of $\frac{\partial }{\partial x^{1}},\frac{%
\partial }{\partial p_{2}}$. The coefficients of these linear
representations are fixed by putting them into the algebra and demanding
that it holds identically. Thus we find 
\begin{eqnarray}
\widehat{x}^{2} &=&i\left( -\vartheta \frac{\partial }{\partial x^{1}}+\frac{%
\partial }{\partial p_{2}}\right) ,  \nonumber \\
\widehat{P}_{1} &=&i\left( -\frac{\partial }{\partial x^{1}}+\omega \frac{%
\partial }{\partial p_{2}}\right) .  \label{24}
\end{eqnarray}
Clearly, for generic values of $\vartheta $ and $\omega ,$ with $\vartheta
\omega \neq 1$, $(\widehat{x}^{2},\widehat{P}_{1})$ behave as independent
operators. Just at the critical point $\omega =1/\vartheta $ the two
operators become proportional, 
\begin{equation}
\widehat{P}_{1}=\omega \widehat{x}^{2}.  \label{25}
\end{equation}
\smallskip The same logic applies if we exchange the role of $(\widehat{x}%
^{1},\widehat{P}_{2})$ with $(\widehat{x}^{2},\widehat{P}_{1})$ and the
result is 
\begin{equation}
\widehat{P}_{2}=-\omega \widehat{x}^{1}.  \label{26}
\end{equation}
Hence we see that, for the critical value $\omega =1/\vartheta ,$ the
operators $\widehat{x}^{a},\widehat{P}_{b}$ satisfying the deformed
Heisenberg algebra must be constrained according to eq.(\ref{19}). In this
case the Hilbert space (representation) of the algebra is no longer defined
by the 2-parametric set of states $|x^{1},p_{2}\rangle $ and degenerates to
a 1-parametric set of states like $|x^{1}\rangle .$

\smallskip

It may be amusing to compare the above phenomenon in reduction of the
dimension of representation to what happens in the case of a short
representation (or BPS\ states) of an ordinary SUSY algebra. In both cases
we deal with an algebra depending on a set of parameters so that on a
special boundary in the space of these parameters the dimension of a generic
representation suddenly decreases. In the case of a short representation,
the parameter space is defined by the mass and charge ($m,q$) parameters and
the boundary is given by the BPS condition, $m=|q|,$ on which the dimension
of a generic representation is reduced by $\frac{1}{2}.$ In our case, the
parameter space consists of $(\vartheta ,\omega )$ in which on the critical
curve, $\omega =1/\vartheta ,$ the basis of the Hilbert space shrinks from a
2-parametric set of states to a 1-parametric set.

\smallskip

\subsection{The Noncommutative Superspace and the Deformed Supertranslation
Algebra}

Our definition of a noncommutative superspace in this paper is a simple
extension of the usual noncommutative space endowed with a set of
anticommuting grassmann coordinates. In other words, we define a
noncommutative superspace as the algebra of the bosonic and fermionic
operators (coordinates) $x^{a},\theta ^{\alpha }$ satisfying 
\begin{equation}
\lbrack x^{a},x^{b}]=i\vartheta \epsilon ^{ab},\qquad \{\theta ^{\alpha
},\theta ^{\beta }\}=0,\qquad [x^{a},\theta ^{\alpha }]=0.  \label{27}
\end{equation}
This corresponds to the algebra of derivatives as follows 
\begin{equation}
\lbrack \frac{\partial }{\partial x^{a}},\frac{\partial }{\partial x^{b}}%
]=-i\omega \epsilon _{ab},\qquad [\frac{\partial }{\partial x^{a}},\frac{%
\partial }{\partial \theta ^{\alpha }}]=0,\qquad \{\frac{\partial }{\partial
\theta ^{\alpha }},\frac{\partial }{\partial \theta ^{\beta }}\}=0.\qquad
\label{28}
\end{equation}
\smallskip We will take $\omega =1/\vartheta $ in all the subsequent
formulae. More generally, we could consider nontrivial (anti)commutation
relations also between $(\theta ,\theta )$ and $(x,\theta )$. However, the
associativity of the $*$-product, which amounts to the satisfaction of the
Jacobi identities, requires trivial relations of these two types, at least
when the noncommutativity parameter ($\vartheta $) is taken a constant \cite
{5.3}.

\smallskip

Let us now assume the same definitions as eqs.(\ref{5}),(\ref{6}) for the
translations and supertranslations \smallskip \smallskip on a noncommutative
superspace. Therefore, the representations of the corresponding generators
are given by the same expressions as in the commutative case (eqs.(\ref{8}%
)), except that in this case the derivative operators $\partial /\partial
x^{a}$ do not commute as in eq.(\ref{28}). Hence, computing the
(anti)commutators of those expressions, taking\smallskip into account the
above fact, we obtain the following modified supertranslation algebra 
\begin{eqnarray}
\lbrack P_{a},P_{b}] &=&i\omega \epsilon _{ab},  \nonumber \\
\lbrack Q_{\alpha },P_{a}] &=&i\omega \epsilon _{ab}(\rho ^{1}\rho
^{b})_{\alpha \beta }\theta ^{\beta },  \nonumber \\
\{Q_{\alpha },Q_{\beta }\} &=&-2(\rho ^{1}\rho ^{a})_{\alpha \beta
}P_{a}+i\omega \epsilon _{ab}(\rho ^{1}\rho ^{a}\theta )_{\alpha }(\rho
^{1}\rho ^{b}\theta )_{\beta }.  \label{29}
\end{eqnarray}
\smallskip As is seen, all the modification terms are proportional to $%
\omega $ and are written in either of the forms $1,\theta ^{+},\theta
^{-},\theta ^{+}\theta ^{-},$ which together form a complete basis on the
space of functions of $(\theta ^{+},\theta ^{-})$. At the first \smallskip
sight, this is not a closed algebra in the sense of ordinary super Lie
algebras. However, we find that we can close this algebra by appending to it
new operators represented by the above combinations. It turns out \cite{11}
that these new operators have the interpretation of the generators of a new
``gauge symmetry'' of the theory, which are represented on a scalar
superfield $S(x,\theta )$ as follows 
\begin{equation}
S\rightarrow f(\theta )S,
\end{equation}
with $f$ being any function of the grassmann odd coordinates. This symmetry
gets mixed with the (super)translation symmetries when the space becomes
noncommutative. Surprisingly, the above algebra does not reduce to the usual
algebra (eq.(\ref{10})) in the limit of zero noncommutativity $\vartheta
\rightarrow 0$, but it does that in the limit of an infinite
noncommutativity $\omega =1/\vartheta $\smallskip $\rightarrow 0$.

\subsection{\protect\smallskip Rotation on a Noncommutative Superspace}

So far we have considered only the (super)translation part of the deformed
super $E^{2}$ algebra. In higher dimensions, the rotational (or Lorentz)
symmetries are generally broken by the spacetime noncommutativity so that
their algebra is not consistently closed with the above deformed translation
algebra\footnote{%
That is the Jacobi identities are not satisfied for the full Poincare
algebra.}. However, in 2 dimensions the $SO(2)$ symmetry is not broken by
the presence of noncommutativity and hence it can be consistently appended
to the deformed translation algebra. This is indeed because the basic
equation $[x^{a},x^{b}]=i\vartheta \epsilon ^{ab}$ is invariant under the
usual $SO(2)$ rotations, 
\begin{equation}
\delta _{\alpha }x^{a}=-\alpha \epsilon ^{ab}x^{b}.  \label{30}
\end{equation}
Denoting the generator of these rotations by $J$, we can write the above
equation in the operator language as 
\begin{equation}
\lbrack J,x^{a}]=-i\epsilon ^{ab}x^{b}.  \label{31}
\end{equation}
\smallskip \smallskip Now, using this equation, the commutator of $J$ with
both sides of the eq.(\ref{17}) becomes

\begin{eqnarray}
\lbrack J,[x^{a},x^{b}]\,] &=&-[x^{b},[J,x^{a}]\,]+[x^{a},[J,x^{b}]\,] 
\nonumber \\
&=&i\epsilon ^{ac}[x^{b},x^{c}]-i\epsilon ^{bc}[x^{a},x^{c}]=0,
\end{eqnarray}
\smallskip showing that the $SO(2)$ rotations are compatible with the
algebra defined by eq. (\ref{17}). \smallskip A \smallskip similar fact can
be proved for the whole algebra defined by the eqs.(\ref{27}), taking into
account that coordinates $\theta ^{\alpha }$ obey the following $SO(2)$
transformations (see also Appendix): 
\begin{equation}
\delta _{\alpha }\theta =i\frac{\alpha }{2}\sigma _{3}\theta .  \label{33}
\end{equation}

\smallskip

\smallskip

Let us now consider the $SO(2)$ variation of an arbitrary function $f(x)$ of
the noncommutative coordinates. More precisely, we are interested in the $%
SO(2)$ variation of the operator $\widehat{f}(\widehat{x})$ which is the
Moyal-Weyl correspondent of the ordinary function $f(x)$. This variation
must be computed using the expansion of $\widehat{f}(\widehat{x}+\delta 
\widehat{x})$, taking into account that \smallskip $\delta \widehat{x}$'s do
not commute with $\widehat{x}$'s, thus preventing the use of the ordinary
Taylor series expansion. Given particular polynomial definitions for $%
\widehat{f}(\widehat{x})$, it is easy to find $\delta \widehat{f}$ by
directly calculating $\widehat{f}(\widehat{x}+\delta \widehat{x})-\widehat{f}%
(\widehat{x})$, and ordering the result in powers of $\delta \widehat{x}$
using the known expressions for $[\widehat{x}^{a},\delta \widehat{x}^{b}]$.
For a general $\widehat{f}(\widehat{x})$, however, it is easier to do the
same task by going first to the function language and then convert the
result into the operator language. For this purpose, we first note that

\begin{equation}
\delta _{\alpha }x^{a}*\partial _{a}f(x)=\delta _{\alpha }x^{a}\partial
_{a}f(x)+\frac{i}{2}\alpha \vartheta \square f(x),  \label{34}
\end{equation}
where $\square \equiv \partial _{a}\partial $\smallskip $_{a}$. Hence, using
the ordinary Taylor expansion of $f(x+\delta x)$ we find 
\begin{equation}
\delta _{\alpha }f(x)=\delta _{\alpha }x^{a}\partial _{a}f(x)=\delta
_{\alpha }x^{a}*\partial _{a}f(x)-\frac{i}{2}\alpha \vartheta \square f(x).
\label{35}
\end{equation}
\smallskip By the Moyal-Weyl map on this equation, we then easily find the
desired expression 
\begin{equation}
\delta _{\alpha }\widehat{f}(\widehat{x})=-\alpha \left( \epsilon _{ab}%
\widehat{x}^{b}\partial _{a}\widehat{f}(\widehat{x})+\frac{i}{2}\vartheta
\square \widehat{f}(\widehat{x})\right) .  \label{36}
\end{equation}
\smallskip Here, the derivative operators are defined through their \textit{%
adjoint} operations, i.e., 
\begin{eqnarray}
\partial _{a}\widehat{f}(\widehat{x}) &=&i\vartheta ^{-1}\epsilon _{ab}[%
\widehat{x}^{b},\widehat{f}(\widehat{x})]\equiv [\widehat{\partial }_{a},%
\widehat{f}(\widehat{x})],  \nonumber \\
\square \widehat{f}(\widehat{x}) &\equiv &[\widehat{\partial }_{a},[\widehat{%
\partial }_{a},\widehat{f}(\widehat{x})]\,].  \label{37}
\end{eqnarray}
\smallskip It is easy to extend the above result to the case of superfields
on a noncommutative superspace. The only change is due to the variation of $%
\theta ^{\alpha }$ giving rise to an additional term in the above expression
as follows 
\begin{equation}
\delta _{\alpha }\widehat{\mathcal{S}}(\widehat{x},\theta )=\alpha \left(
-\epsilon _{ab}\widehat{x}^{b}\partial _{a}-\frac{i}{2}\vartheta \square +%
\frac{i}{2}\sigma _{\alpha \beta }^{3}\theta ^{\beta }\frac{\partial }{%
\partial \theta ^{\alpha }}\right) \widehat{\mathcal{S}}(\widehat{x},\theta
).  \label{38}
\end{equation}
The superspace representation of the $SO(2)$ generator $J$ is read from this
expression to be 
\begin{equation}
-J=L+S,  \label{39}
\end{equation}
where 
\begin{eqnarray}
L &\equiv &-i\epsilon _{ab}x^{a}\partial _{b}-\frac{1}{2}\omega ^{-1}\square
,  \nonumber \\
S &\equiv &\frac{1}{2}\sigma _{\alpha \beta }^{3}\theta ^{\alpha }\frac{%
\partial }{\partial \theta ^{\beta }}.  \label{40}
\end{eqnarray}
This $L,\,S$ specify the \textit{orbital }and \textit{spin} angular momentum%
\textit{\ }operators, respectively. Compared to the commutative expressions,
noncommutativity modifies only the orbital part of $J$ but not its spin part
as a result of non-deforming the algebra of the odd coordinates. Now, using
the relation (\ref{19}) between $x,P$, or alternatively $\partial
_{a}=i\omega \epsilon _{ab}x^{b}$, we find that $L$ is written in either of
the following forms: 
\begin{equation}
L=\frac{1}{2}(x^{1}P_{2}-x^{2}P_{1})=-\frac{1}{2\omega }%
(P_{1}^{2}+P_{2}^{2}).  \label{41}
\end{equation}
The first expression is just $1/2$ the classical (commutative) expression of 
$L$. It turns out that this factor is crucial in recovering the correct
commutation relation between $L,P_{a}$, i.e. 
\begin{equation}
\lbrack L,P_{a}]=-i\epsilon _{ab}P_{b}\,,
\end{equation}
as is required by the $SO(2)$ transformation of $P_{a}\,$as a vector.
\smallskip The second expression in (\ref{41}) shows that the orbital the
angular momentum of a particle on a noncommutative plane is proportional to
its (Euclidean) mass squared $M^{2}\equiv (P_{1})^{2}+(P_{2})^{2}$ ; i.e. 
\begin{equation}
L=-\frac{1}{2\omega }M^{2}.  \label{42}
\end{equation}
\smallskip This is comparable to the expression in \cite{7.7} for the
generator of rotations on the phase space of a particle in 2D. (see also 
\cite{7.1})

\smallskip

\subsection{The Complete Super $E_{\vartheta }^{2}$ Algebra}

We are now ready to obtain the full \smallskip super $E_{\vartheta }^{2}$
algebra by computing the (anti)commutators of all pairs of its generators $%
P_{a},Q_{\alpha },J$ from their explicit representations by eqs.(\ref{8}), (%
\ref{39}), (\ref{40}). Besides that, we introduce the new operators $%
O_{+},O_{-},T$ via their representations (as noted below eq.(\ref{29})) as
follows 
\begin{equation}
O_{+}\equiv \theta ^{-},\qquad O_{-}\equiv \theta ^{+},\qquad T\equiv \theta
^{+}\theta ^{-}.
\end{equation}
\smallskip Further we define the complex momenta as 
\begin{equation}
P_{\pm }\equiv P_{1}\mp iP_{2}
\end{equation}
It is now easy to show that the algebra of the generators $P_{\pm },Q_{\pm
},J,O_{\pm },T$ is a closed algebra whose non-trivial relations are
summarized in the following 
\begin{eqnarray}
\lbrack P_{+},P_{-}] &=&2\omega ,\quad \qquad \{Q_{+},Q_{-}\}=2\omega T,\quad
\nonumber \\
\lbrack Q_{+},P_{-}] &=&-2\omega O_{-},\qquad [Q_{-},P_{+}]=2\omega O_{+}, 
\nonumber \\
(Q_{+})^{2} &=&-P_{+},\quad \qquad (Q_{-})^{2}=-P_{-},  \nonumber \\
\{Q_{+},O_{-}\} &=&1,\quad \qquad \{Q_{-},O_{+}\}=1,  \nonumber \\
\lbrack Q_{+},T] &=&O_{+},\quad \qquad [Q_{-},T]=-O_{-},  \nonumber \\
\lbrack J,P_{+}] &=&P_{+},\quad \qquad [J,P_{-}]=-P_{-},  \nonumber \\
\lbrack J,Q_{+}] &=&\frac{1}{2}Q_{+},\qquad \quad [J,Q_{-}]=-\frac{1}{2}%
Q_{-},  \nonumber \\
\lbrack J,O_{+}] &=&\frac{1}{2}O_{+},\quad \qquad [J,O_{-}]=-\frac{1}{2}%
O_{-}.  \label{43}
\end{eqnarray}

\smallskip This is a consistent super Lie algebra in the sense that the
coefficients on its RHS have been set in a way that it automatically
satisfies the Jacobi identities. For example, using this algebra, one can
easily verify the following identities 
\begin{eqnarray}
\lbrack
Q_{+},\{Q_{+},Q_{-}\}]+[Q_{+},\{Q_{-},Q_{+}\}]+[Q_{-},\{Q_{+},Q_{+}\}] &=& 
\nonumber \\
2\omega O_{+}+2\omega O_{+}-4\omega O_{+} &=&0,  \nonumber \\
\lbrack
P_{+},\{Q_{-},Q_{-}\}]+\{Q_{-},[Q_{-},P_{+}]\}+\{Q_{-},[Q_{-},P_{+}]\,\} &=&
\nonumber \\
-4\omega +2\omega +2\omega &=&0,  \nonumber \\
\lbrack J,\{Q_{+},Q_{+}\}]+\{Q_{+},[Q_{+},J]\}+\{Q_{+},[Q_{+},J]\,\} &=&_{{}}
\nonumber \\
-2P_{+}+P_{+}+P_{+} &=&0,  \nonumber \\
\lbrack J,\{Q_{+},Q_{-}\}]+\{Q_{+},[Q_{-},J]\}+\{Q_{-},[Q_{+},J]\,\} &=& 
\nonumber \\
0+\omega T-\omega T &=&0,  \nonumber \\
\lbrack J,[Q_{+},T]\,]+[Q_{+},[T,J]\,]+[J,[Q_{+},T]\,] &=&  \nonumber \\
O_{+}+0-O_{+} &=&0.  \label{44}
\end{eqnarray}
\smallskip \smallskip This is not a surprising feature of the algebra as it
was derived using explicit representations for its generators on a Hilbert
space consisting of all the superfields on the given (noncommutative)
superspace. Such an algebra must necessarily satisfy the Jacobi identities
as a result of its \textit{associativity.}

As is clear, the modifications due to the noncommutativity of space appear 
\textit{only} in the $P,Q$ sector of the above algebra. More significantly,
all the noncommutativity modifications, which are proportional to $1,O_{\pm
},T$, commute with any (operator valued) superfield $\widehat{S}(\widehat{x}%
,\theta )$, since they do not depend on $\widehat{x}^{a}$ at all. Due to
this feature, these modifications are not observable in the algebra of the
SUSY variations in a NCFT which is a Moyal deformation of a usual SUSY field
theory (see next section).

Finally, we note that the operators $Q_{\pm }$ ($P_{\pm }$) behave as the
ladder operators of $J$ shifting its eigenvalue by $\pm 1/2\,(\pm 1)$ units.

\section{\protect\smallskip Spectrum of the Super\thinspace $E_{\vartheta
}^{2}$ Algebra}

\subsection{The Purely Bosonic Case}

\smallskip Let us begin the analysis of the spectrum by considering first
the spectrum of the purely bosonic translation subalgebra,

\begin{equation}
\lbrack P_{+},P_{-}]=2\omega .  \label{45}
\end{equation}
Noting to $(P_{+})^{\dagger }=P_{-},$ this is reduced to a simple harmonic
oscillator algebra by taking 
\[
P_{+}=\sqrt{2\omega }a,\qquad P_{-}=\sqrt{2\omega }a^{\dagger }, 
\]
\begin{equation}
\lbrack a,a^{\dagger }]=1.  \label{46}
\end{equation}
The Euclidean mass squared operator in two dimensions, 
\begin{equation}
M^{2}=P_{1}^{2}+P_{2}^{2}=P_{+}P_{-}-\omega =P_{-}P_{+}+\omega ,  \label{47}
\end{equation}
\smallskip is proportional to the Hamiltonian of the harmonic oscillator,
i.e. 
\begin{equation}
M^{2}=2\omega \left( a^{\dagger }a+\frac{1}{2}\right) .  \label{48}
\end{equation}
So, the representation space of the translation algebra in two dimensions is
\smallskip spanned by the energy eigenstates of a harmonic oscillator in one
dimensions, which are the same as the mass eigenstates in two dimensions;
i.e. 
\begin{equation}
\mathcal{H}=\mathrm{span}\{|n\rangle ;\quad (a^{\dagger }a)|n\rangle
=n|n\rangle ;\quad n=0,1,2,...\}.  \label{49}
\end{equation}
The corresponding $M^{2}$ eigenvalues are 
\begin{equation}
m_{n}^{2}=2\omega \left( n+\frac{1}{2}\right) ,  \label{50}
\end{equation}
which by eq.(\ref{42}) correspond to the quantized values of $L$ as follows 
\begin{equation}
l_{n}=-\left( n+\frac{1}{2}\right) .
\end{equation}
The effect of $P_{\pm }$\smallskip on this spectrum of states is simply to
higher or lower the value of $n$ according to 
\begin{equation}
a|n\rangle =\sqrt{n}|n-1\rangle ,\quad \quad a^{\dagger }|n\rangle =\sqrt{n+1%
}|n+1\rangle .  \label{51}
\end{equation}
This is indeed a basis for the \textit{fundamental} representation of the
noncommutative translation algebra. Alternatively, we can construct an 
\textit{adjoint }representation by forming a basis on the space of the
operators in terms of the states $|n\rangle $ 
\begin{equation}
\mathcal{H}^{*}=\mathrm{span}\{|m\rangle \langle n|;\quad m,n=0,1,2,...\}.
\label{52}
\end{equation}
A \textit{unitary }transformation $U=\mathrm{e}^{iA}$ corresponding to a
symmetry operation in the completed algebra acts on $\mathcal{H}$ and $%
\mathcal{H}^{*}$ as follows 
\begin{eqnarray}
|n\rangle &\longrightarrow &U|n\rangle ,  \nonumber \\
|m\rangle \langle n| &\longrightarrow &U|m\rangle \langle n|U^{\dagger }.
\label{53}
\end{eqnarray}
Now, since any field $\widehat{\phi }$\smallskip $(\widehat{x})$ in the
operator formulation has an expansion in the basis defined in eq.(\ref{52}), 
\begin{equation}
\widehat{\phi }\smallskip (\widehat{x})=\sum_{m,n=0}^{\infty
}c_{mn}|m\rangle \langle n|,  \label{54}
\end{equation}
the above unitary transformation must change $\widehat{\phi }\smallskip (%
\widehat{x})$ as 
\begin{equation}
\widehat{\phi }\smallskip (\widehat{x})\rightarrow U\widehat{\phi }%
\smallskip (\widehat{x})U^{\dagger }.  \label{55}
\end{equation}
For an infinitesimal transformation of this kind, the change of a field in
NCFT in either of its operator and ordinary formulations becomes 
\begin{equation}
\delta _{A}\widehat{\phi }\smallskip (\widehat{x})=i[\widehat{A}\smallskip ,%
\widehat{\phi }\smallskip (\widehat{x})]\quad \leftrightarrow \quad \delta
_{A}\phi (x)=i[A(x),\phi (x)]_{*}\,.  \label{56}
\end{equation}
Here, the operators $\widehat{A}\smallskip ,\widehat{\phi }\smallskip (%
\widehat{x})$ and the functions $A(x),\phi (x)$ are related by the
Moyal-Weyl map. This shows that the noncommutative fields always transform
in the adjoint representation of the noncommutative supergroup. A direct
consequence of this fact is that it is not possible to observe the
noncommutativity modifications of the superalgebra in a NCFT which is
obtained by Moyal deformation of an ordinary SUSY field theory \cite{6.4}.
This is because all the modifications found in eqs.(\ref{43}), being pure
functions of $\theta ^{\alpha }$, commute with any superfield $\widehat{%
\mathcal{S}}(\widehat{x},\theta )$. As a result, these modifications are
removed from all the commutators, if one tries to recover the superalgebra
through the algebra of the variations of the superfields (e.g. $[\delta
_{A},\delta _{B}]\widehat{\mathcal{S}}$ for each pair of the generators $A,B$%
), and hence, the superalgebra is apparently unchanged. This is compatible
with the result of \cite{6.4} found in the context of the noncommutative WZ
model using explicit expressions for its conserved (super)charges.

\smallskip

\subsection{\protect\smallskip The Supersymmetric Case}

\textit{The commutative limit:}

Before completing the spectrum of \smallskip the super $E_{\vartheta }^{2}$
algebra, it is worthwhile to have a look at its commutative counterpart
which is obtained by going to the $\omega \rightarrow 0$ limit. More
precisely, when $\omega =0$ in eqs.(\ref{43}), we obtain two separated
subalgebras, one for $P_{a},Q_{\alpha },O_{\alpha },T$ and the other for $%
P_{a},Q_{\alpha },J$, which the latter defines the commutative super $E^{2}$%
algebra through the non-vanishing (anti)commutators: 
\begin{eqnarray}
Q_{+}^{2} &=&-P_{+},\quad Q_{-}^{2}=-P_{-},\quad  \nonumber \\
\lbrack J,P_{\pm }] &=&\pm P_{\pm },\quad [J,Q_{\pm }]=\pm \frac{1}{2}Q_{\pm
}\,.  \label{57}
\end{eqnarray}
\smallskip This algebra has a single Casimir operator which is defined by
the mass squared: 
\begin{equation}
M^{2}=P_{+}P_{-}\,.  \label{58}
\end{equation}
\smallskip \smallskip This is an $SO(2)$-invariant quantity and hence
commutes with $J$, 
\begin{equation}
\lbrack M^{2},J]=0.  \label{59}
\end{equation}
\smallskip Indeed, $(M^{2},J)\,$form the maximal commuting set of operators
for the super $E^{2}$ algebra and so an \textit{irreducible }representation
of this algebra is spanned by the simultaneous eigenstates of these
operators: 
\begin{eqnarray}
M^{2}|m,j\rangle &=&m^{2}|m,j\rangle ,  \nonumber \\
J|m,j\rangle &=&j|m,j\rangle .  \label{60}
\end{eqnarray}
\smallskip \smallskip Here, the mass squared eigenvalue $m^{2}$ is any 
\textit{positive} or \textit{negative} real number, $\,$provided we do not
assume any restriction on the positive-definiteness of the norm in the
Hilbert space. On the other hand, the angular momentum eigenvalue $j$ in 2D
(despite in higher dimensions) is not restricted to half-integers, but it
takes \textit{continuous} real values. However, because of the $\pm 1/2$
shifts of $j$ due to the action of $Q\pm $, any irreducible representation
of the algebra consists only of those states whose values of $j$ differ by
the half integers. This means that different irreducible representations of
the super $E^{2}$algebra (for a fixed $m^{2}$) are distinguished by a real
number $\nu $, $0\leq \nu $ $<\frac{1}{2}$, which restricts the values of $j$
to $j=\nu +\frac{n}{2}$, with $n\in \Bbb{Z}$. We note that, for a fixed $%
m^{2}$, the states $|m,j\rangle $ for different $j$ 's are orthogonal since $%
J$ is a Hermitian operator. \smallskip

\smallskip

The action of $Q_{\pm }$ on these states is specified as follows 
\begin{equation}
Q_{\pm }|m,j\rangle =C_{\pm }(m,j)|m,j\pm \frac{1}{2}\rangle ,  \label{61}
\end{equation}
\smallskip where $C_{\pm }(m,j)$ are coefficients to be determined. The
action of $P_{\pm }$ in this representation is then specified \smallskip by
simply acting twice by $Q_{\pm }$ which using the above equation becomes 
\begin{equation}
P_{\pm }|m,j\rangle =-C_{\pm }(m,j)C_{\pm }(m,j\pm \frac{1}{2})|m,j\pm
1\rangle .  \label{62}
\end{equation}
\smallskip Taking the inner product of eqs.(\ref{61}),(\ref{62}) with
another arbitrary state of representation, $|m,j^{\prime }\rangle ,$ and
also using the norms of the same equations, we obtain a set of equations
involving the constants $C(m,j)$ and the norms of states $\langle
m,j|m,j\rangle $ for different $j$'s. These can be analyzed in detail and
give the following results:

\smallskip

a) For $m^{2}>0$, all the states $|m,j\rangle $ in the representation are 
\textit{null} and, with a suitable choice of their normalizations, we can
write 
\begin{eqnarray}
C_{+}(m,j) &=&-\sqrt{m},\quad C_{-}(m,j)=i(-1)^{[2j]}\sqrt{m},  \nonumber \\
\langle m,j|m,j\rangle &=&0.  \label{63}
\end{eqnarray}
Here $[x]$ denotes the integer part of the real number $x.$

b) For $m^{2}=0$ , we have the following restrictions between $C_{\pm }$ and
the norm of the states 
\begin{eqnarray}
C_{+}(0,j)C_{-}(0,j+1/2) &=&0,  \nonumber \\
C(0,j\pm 1/2)\langle 0,j|0,j\rangle &=&0,  \label{64}
\end{eqnarray}
which allows for both the positive and negative normed states, as well as
the null states.

c) For $m^{2}=-\mu ^{2}<0$ , we have both positive and negative normed
states in the representation and, with a suitable normalization, we can
write 
\begin{eqnarray}
C_{+}(\mu ,j) &=&-\sqrt{\mu },\quad C_{-}(\mu ,j)=(-1)^{[2j]}\sqrt{\mu }, 
\nonumber \\
\langle \mu ,j|\mu ,j\rangle &=&(-1)^{\frac{1}{2}[2j]([2j]-1)}.  \label{65}
\end{eqnarray}
\smallskip \smallskip We can construct \textit{reducible }representations of
the algebra out of the tensor products of two of its irreducible
representations. For example, in the superfield representation on a
Euclidean $\Bbb{R}^{2}$plane, we indeed use \smallskip the tensor product of
a representation with $j\in \Bbb{Z}$ but with an arbitrary $m^{2},$ and
another one with $j=0,\pm \frac{1}{2}$ and $m^{2}=0,$ corresponding to the
functions of the form $f(x)$ and $g(\theta )$, respectively.\smallskip Other
geometries may also be handled by using other representations of the same
algebra. For example, with $j\notin \Bbb{Z}$ in the first representation, we
could obtain a superfield representation on a conic geometry rather than on
the Euclidean plane.

\smallskip

\smallskip

\smallskip

\textit{The noncommutative case:}

With insights from the commutative limit, it must now be evident that
finding the spectrum of the complete super $E_{\vartheta }^{2}$ algebra is
equivalent to finding a suitable basis for the expansion of a noncommutative
superfield. Any such superfield is expressed, via a Moyal-Weyl map, as an
operator, 
\begin{equation}
\widehat{\mathcal{S}}\smallskip (\widehat{x},\theta )=\widehat{\phi }%
\smallskip (\widehat{x})+\theta ^{+}\widehat{\psi }_{+}(\widehat{x})+\theta
^{-}\widehat{\psi }_{-}(\widehat{x})+\theta ^{+}\theta ^{-}\widehat{F}%
\smallskip (\widehat{x}),  \label{66}
\end{equation}
whose components ($\widehat{\phi }\smallskip ,\,\widehat{F},\widehat{\psi }%
_{\pm }$) are themselves a set of (bosonic or fermionic) operators. Taking
the basis for functions of $\widehat{x}$ as in eq.(\ref{52}) and that of
functions of $\theta $ as 
\begin{equation}
\mathcal{K}=\mathrm{span}\{1,\theta ^{+},\theta ^{-},\theta ^{+}\theta
^{-}\},  \label{67}
\end{equation}
the representation space of our algebra in the adjoint case is simply
spanned by the tensor product 
\begin{equation}
\mathcal{L}^{*}=\mathcal{H}^{*}\otimes \mathcal{K}.  \label{68}
\end{equation}
The corresponding basis for $\widehat{\mathcal{S}}\smallskip (\widehat{x}%
,\theta )$ is thus of the form: $|m,n\rangle |i\rangle $ ($m,n=0,1,2,....$
and $i=1,...,4$), where $|i\rangle \in \mathcal{K}$ is one of the four
states represented by $1,\theta ^{+},\theta ^{-},\theta ^{+}\theta ^{-}$.
Alternatively, we could construct the fundamental representation by taking
the tensor product 
\begin{equation}
\mathcal{L}=\mathcal{H}\otimes \mathcal{K},
\end{equation}
\smallskip whose basis vectors consist of the states of the form $|n\rangle
|i\rangle $.

It is useful to organize the states in $\mathcal{K}$ according to the
eigenvalues of the two commuting operators $R,S$ defined as follows 
\begin{eqnarray}
R &\equiv &\frac{1}{2}\left( \theta ^{+}\frac{\partial }{\partial \theta ^{+}%
}+\theta ^{-}\frac{\partial }{\partial \theta ^{-}}\right) ,  \nonumber \\
S &\equiv &\frac{1}{2}\left( \theta ^{+}\frac{\partial }{\partial \theta ^{+}%
}-\theta ^{-}\frac{\partial }{\partial \theta ^{-}}\right) .  \label{70}
\end{eqnarray}
The operator $S$ here is the same as the spin operator defined by eq.(\ref
{40}). Indeed, the eigenvalues of $S$ specify the spins of the states in $%
\mathcal{K}$ while those of $R$ specify their length dimensions. It is easy
to see that these operators satisfy the following equations 
\begin{eqnarray}
2R^{3}-3R^{2}+R &=&0,  \nonumber \\
4S^{3}-S &=&0.  \label{71}
\end{eqnarray}
\smallskip showing that $R$ and $S$ have the eigenvalues $r=0,\frac{1}{2},1$
and $s=-\frac{1}{2},0,\frac{1}{2}$, respectively. One can then easily check
that the basis vectors in $\mathcal{K}$ are the simultaneous eigenstates of $%
R,S$ , which henceforth are identified with $|r,s\rangle $ as follows 
\begin{eqnarray}
|\frac{1}{2},\frac{1}{2}\rangle &\equiv &(2\omega )^{-\frac{1}{4}}\theta
^{+},\quad |\frac{1}{2},-\frac{1}{2}\rangle \equiv (2\omega )^{-\frac{1}{4}%
}\theta ^{-},  \nonumber \\
|0,0\rangle &\equiv &1,\quad \quad \quad |1,0\rangle \equiv (2\omega )^{-%
\frac{1}{2}}\theta ^{+}\theta ^{-}.  \label{72}
\end{eqnarray}
\smallskip and all other combinations of $r,s$ are identified with the zero
state. It is then easy to express the operations of $\theta ^{\alpha
},\partial /\partial \theta ^{\alpha }$ on each of the states $|r,s\rangle $
in terms of some other $|r,s\rangle $.

\smallskip

\smallskip An arbitrary state $|f\rangle $ in $\mathcal{K}$ is expressed as
a function $f(\theta )$ of the Grassmann coordinates. A natural definition
of the inner product in $\mathcal{K}$ for two states $|f\rangle ,|g\rangle $
is given by 
\begin{equation}
\langle f|g\rangle =\int d^{2}\theta \;f^{*}(\theta )g(\theta ).  \label{73}
\end{equation}
Using this definition, one can check that the chosen basis for $\mathcal{K}$
is orthogonal. On the other hand, defining the norm in $\mathcal{K}$ as
usual by $||f||^{2}\equiv \langle f|f\rangle $, one finds that $\mathcal{K}$
contains both positive and negative normed states as well as null states.

\smallskip Now that the effects of $(a,a^{\dagger })$ on the states in $%
\mathcal{H\,}$or $\mathcal{H}^{*}$ and that of $(\theta ^{\alpha },\partial
/\partial \theta ^{\alpha })$ on the states in $\mathcal{K}$ are well
understood, the problem of representation of the super $E^{2}$ algebra
reduces to expressing all its generators in terms of $(a,a^{\dagger },\theta
^{\alpha },\partial /\partial \theta ^{\alpha })$. Using the previous
representations of these generators, thus we find 
\begin{eqnarray}
P_{+} &=&\sqrt{2\omega }a,\quad \quad \quad \quad P_{-}=\sqrt{2\omega }%
a^{\dagger },  \nonumber \\
Q_{+} &=&\frac{\partial }{\partial \theta ^{-}}-\sqrt{2\omega }\theta
^{-}a,\quad Q_{-}=\frac{\partial }{\partial \theta ^{+}}-\sqrt{2\omega }%
\theta ^{+}a,  \label{74} \\
O_{+} &=&\theta ^{-},\qquad O_{-}=\theta ^{+},\qquad T=\theta ^{+}\theta
^{-}, \\
J &=&\left( a^{\dagger }a+\frac{1}{2}\right) -\frac{1}{2}\left( \theta ^{+}%
\frac{\partial }{\partial \theta ^{+}}-\theta ^{-}\frac{\partial }{\partial
\theta ^{-}}\right) .
\end{eqnarray}
\smallskip For the states in the fundamental representation, $|n\rangle
|r,s\rangle $, these generators act as operators from the left, while for
the states in the adjoint representation, $|m,n\rangle |r,s\rangle $, the
generators act via taking their (anti)commutators with these states\footnote{%
Recall that the states in the adjiont representation $\mathcal{L}^{*}$ are
in the form of operators acting on the states in the fundamental
representation $\mathcal{L}.$}. As such, we obtain the two representations
of the algebra (\ref{43}) as follows

\smallskip

$_{{}}$\smallskip

\smallskip

\smallskip

\textit{The Fundamental Representation:} 
\begin{eqnarray}
P_{+}|n\rangle |r,s\rangle &=&(2\omega )^{\frac{1}{2}}\sqrt{n}|n-1\rangle
|r,s\rangle ,  \nonumber \\
P_{-}|n\rangle |r,s\rangle &=&(2\omega )^{\frac{1}{2}}\sqrt{n+1}|n+1\rangle
|r,s\rangle ,  \nonumber \\
Q_{+}|n\rangle |r,s\rangle &=&(-)^{r+s}(2\omega )^{\frac{1}{4}}\left(
|n\rangle |r+\frac{1}{2},s-\frac{1}{2}\rangle -\sqrt{n}|n-1\rangle |r-\frac{1%
}{2},s+\frac{1}{2}\rangle \right) ,  \nonumber \\
\quad \quad \quad Q_{-}|n\rangle |r,s\rangle &=&(2\omega )^{\frac{1}{4}%
}\left( |n\rangle |r-\frac{1}{2},s-\frac{1}{2}\rangle -\sqrt{n+1}|n+1\rangle
|r+\frac{1}{2},s+\frac{1}{2}\rangle \right) ,  \nonumber \\
J|n\rangle |r,s\rangle &=&\left( n-s+\frac{1}{2}\right) |n\rangle
|r,s\rangle ,  \nonumber \\
O_{+}|n\rangle |r,s\rangle &=&(2\omega )^{-\frac{1}{4}}|n\rangle |r+\frac{1}{%
2},s+\frac{1}{2}\rangle ,  \nonumber \\
O_{-}|n\rangle |r,s\rangle &=&(-)^{r+s}(2\omega )^{-\frac{1}{4}}|n\rangle |r+%
\frac{1}{2},s-\frac{1}{2}\rangle ,  \nonumber \\
T|n\rangle |r,s\rangle &=&(-)^{r+s+1}(2\omega )^{-\frac{1}{2}}|n\rangle
|r+1,s\rangle .  \label{75}
\end{eqnarray}

\smallskip

\smallskip

\smallskip

\textit{The Adjoint Representation:} 
\begin{eqnarray}
P_{+}|m,n\rangle |r,s\rangle &=&(2\omega )^{\frac{1}{2}}\left( \sqrt{m}%
|m-1,n\rangle |r,s\rangle -\sqrt{n+1}|m,n+1\rangle |r,s\rangle \right) , 
\nonumber \\
P_{-}|m,n\rangle |r,s\rangle &=&(2\omega )^{\frac{1}{2}}\left( \sqrt{m+1}%
|m+1,n\rangle |r,s\rangle -\sqrt{n}|m,n-1\rangle |r,s\rangle \right) , 
\nonumber \\
Q_{+}|m,n\rangle |r,s\rangle &=&(-)^{r+s}(2\omega )^{\frac{1}{4}}\left( 
\begin{array}{c}
|m,n\rangle |r-\frac{1}{2},s+\frac{1}{2}\rangle \\ 
-\sqrt{m}|m-1,n\rangle |r+\frac{1}{2},s-\frac{1}{2}\rangle \\ 
+\sqrt{n+1}|m,n+1\rangle |r+\frac{1}{2},s-\frac{1}{2}\rangle
\end{array}
\right) ,  \nonumber \\
\quad \quad \quad Q_{-}|m,n\rangle |r,s\rangle &=&(2\omega )^{\frac{1}{4}%
}\left( 
\begin{array}{c}
|m,n\rangle |r-\frac{1}{2},s-\frac{1}{2}\rangle \\ 
-\sqrt{m+1}|m+1,n\rangle |r+\frac{1}{2},s+\frac{1}{2}\rangle \\ 
+\sqrt{n}|m,n-1\rangle |r+\frac{1}{2},s+\frac{1}{2}\rangle
\end{array}
\right) ,  \nonumber \\
J|m,n\rangle |r,s\rangle &=&\left( m-n-s\right) |m,n\rangle |r,s\rangle , 
\nonumber \\
O_{+}|m,n\rangle |r,s\rangle &=&O_{-}|m,n\rangle |r,s\rangle =T|m,n\rangle
|r,s\rangle =0.  \label{76}
\end{eqnarray}

\smallskip

\smallskip

As we see, despite in the commutative case, here the angular momentum
parameter is not arbitrary but restricted to integer and half integer values
in both representations. In the fundamental case it equals to 
\begin{equation}
j=n-s+\frac{1}{2},  \label{77}
\end{equation}
while in the adjoint case it is given by 
\begin{equation}
j=m-n-s.  \label{78}
\end{equation}
In both cases one can find $SO(2)$ invariant states solving the $j=0$
condition. In particular, in the adjoint representation, we find the $s=0,$ $%
m=n$ solution corresponding to $|n\rangle \langle n|$ states, which form a
complete basis of radially symmetric configurations for a scalar field on a
noncommutative plane \cite{10}. The last line in eq.(\ref{76}) is another
indication that the states of the adjoint representation (i.e. the
noncommutative fields) do not ``feel'' the effect of the new transformations
generated by $O_{\pm },\,T$ . This is why the modifications of eqs.(\ref{43}%
) do not appear in the algebra of a supersymmetric NCFT.

\smallskip

\smallskip

\section{Conclusion}

We can summarize the main results of this paper as follows:

\smallskip

$1)$ We found a noncommutative generalization (deformation) of the standard
super $E^{2}$ algebra with $\mathcal{N}$ $=1$ SUSY. The generalized algebra
involves new generators which constitute the basis of the grassmann algebra
in 2 dimensions.

\smallskip

$2)$ $\smallskip $The construction was based on a constraint relating the
momenta to the coordinates of a particle on a noncommutative plane. This
could be interpreted, as in the noncommutative Landau problem, as working in
the regime with the critical value of the magnetic field $\Phi =-\Theta
^{-1} $.

\smallskip

$3)$ The deformed algebra goes back to its non-deformed (commutative)
version in the limit of infinite noncommutativity ($\vartheta \rightarrow
\infty $).

\smallskip

$4)$ We found the spectrum of the deformed superalgebra both in its
fundamental and adjoint representations. The spectrum essentially consists
of the tensor product of the mass eigenstates with the spin eigenstates.

\smallskip

$5)$ The mass and orbital angular momentum operators were found to be
related as $L=-\frac{1}{2}\vartheta M^{2}$. Both the mass and angular
momentum eigenvalues are quantized. Despite in the commutative case, the
mass operator is not a Casimir of the algebra and as a result the
(super)translation generators mix several mass eigenstates.

\smallskip

\smallskip A generic feature of the deformation found in this paper (which
can be easily generalized to other superalgebras in higher dimensions) is
that, despite in the commutative spaces, several SUSY generators have
non-vanishing (anti)commutators with themselves and with the translation
generators. Further, these non-vanishing (anti)commutators are not of the
same type as the (super)translation generators, but they involve new types
of generators. This is in contrast to another type of deformation found in 
\cite{4.1} for the case of a constant $H$-field background, which also have
similar non-vanishing (anti)commutators, but of course without any new
generators.

\smallskip

We pointed out that the modifications of the superalgebra of the type that
were found in this paper are not observable in the NCFT's which are Moyal
deformations of the ordinary SUSY field theories. This is due to the fact
that the fields in such theories transform in the adjoint representation of
the supergroup and they do not carry the ``charges'' of the new generators
in the algebra. Alternatively, the modifications become observable in a
field theory whose field content transform in the fundamental representation
of the supergroup. In that case the additional generators would be
interpreted as the generators of some new symmetries of the underlying
theory \cite{11}.

\smallskip

\smallskip

\smallskip

{\large Acknowledgments}

The author would like to thank S. Parvizi and M.R. Garousi for helpful
discussions.

\smallskip

{\LARGE \smallskip }

\smallskip

\section{Appendix: Spinors in 2 Dimensions}

\renewcommand{\theequation}{A.\arabic{equation}}\setcounter{equation}{0}In
this appendix we present some of the properties of 2D spinors that are
necessary for fixing our notations and conventions throughout this paper. We
work with a flat spacetime with the Cartesian coordinates $x^{a}$, which is
endowed with the Euclidean metric $\delta _{ab}$ ($a,b=1,2$). The isometry
group of this space is the Euclidean group $E^{2}$ consisting of the $SO(2)$
rotations and 2D translations. Dirac spinors in 2D have $2^{[D/2]}=2$
complex components and they can be reduced to minimal spinors with 2 real
components or, alternatively, 2 complex components conjugate to each other.
A real spinorial representation of the $SO(2)$ group is defined as a two
component object $\theta ,$ which under the $SO(2)$ rotations 
\begin{equation}
\left( 
\begin{array}{l}
x^{1} \\ 
x^{2}
\end{array}
\right) \rightarrow \left( 
\begin{array}{ll}
\cos \alpha & -\sin \alpha \\ 
\sin \alpha & \cos \alpha
\end{array}
\right) \left( 
\begin{array}{l}
x^{1} \\ 
x^{2}
\end{array}
\right) ,  \label{A1}
\end{equation}
it is transformed as follows 
\begin{equation}
\left( 
\begin{array}{l}
\theta ^{1} \\ 
\theta ^{2}
\end{array}
\right) \rightarrow \left( 
\begin{array}{ll}
\cos \frac{\alpha }{2} & -\sin \frac{\alpha }{2} \\ 
\sin \frac{\alpha }{2} & \cos \frac{\alpha }{2}
\end{array}
\right) \left( 
\begin{array}{l}
\theta ^{1} \\ 
\theta ^{2}
\end{array}
\right) .  \label{A2}
\end{equation}
\smallskip \smallskip From a suitable combination of the bilinear products $%
\theta ^{\alpha }\theta ^{\beta }$ one can form $SO(2)$ scalars and vectors.
Defining the complex coordinate $z\equiv x^{1}+ix^{2}$ and the complex
spinor $\theta $ with its components $\theta ^{\pm }\equiv \theta ^{1}\pm
i\theta ^{2}$ (hence $(\theta ^{+})^{*}=\theta ^{-}$), one can rewrite the
above $SO(2)$ transformations as follows 
\begin{eqnarray}
z &\rightarrow &e^{i\alpha }z,  \nonumber \\
\theta &\rightarrow &e^{i\alpha \sigma _{3}/2}\theta .  \label{A3}
\end{eqnarray}
\smallskip \smallskip We recall that the spinorial degrees of freedom are
taken as grassmann anticommuting numbers throughout this paper. It is easy
then to form bilinear invariants out of two grassmann spinors $\theta ,\psi $
in the above complex representation. This gives 
\begin{eqnarray}
\text{scalar} &:&\quad \overline{\theta }\psi ,  \nonumber \\
\text{pseudo scalar} &:&\quad i\overline{\theta }\sigma ^{3}\psi ,  \nonumber
\\
\text{vector} &:&\quad \overline{\theta }\rho ^{a}\psi .  \label{A4}
\end{eqnarray}
where $\overline{\theta }$ is defined as 
\begin{equation}
\overline{\theta }\equiv -\theta ^{\dagger }\sigma ^{3},  \label{A5}
\end{equation}
\smallskip and $\rho ^{a}$ are Dirac gamma matrices in 2 Euclidean
dimensions which satisfy the Clifford algebra of $SO(2),$%
\begin{equation}
\{\rho ^{a},\rho ^{b}\}=2\delta ^{ab}.  \label{A6}
\end{equation}
We use the following representation of these matrices 
\begin{equation}
\rho ^{1}=\sigma ^{2},\qquad \rho ^{2}=\sigma ^{1}.  \label{A7}
\end{equation}
In these formulae, $\sigma ^{1},\sigma ^{2},\sigma ^{3}$ denote the standard
Pauli matrices. It is obvious that in this representation $(\rho
^{a})^{\dagger }$ $=\rho ^{a}$. We can easily show that the following
properties hold 
\begin{eqnarray}
(\overline{\theta }\psi )^{*} &=&\overline{\theta }\psi =-\overline{\psi }%
\theta ,  \nonumber \\
(\overline{\theta }\sigma ^{3}\psi )^{*} &=&-\overline{\theta }\sigma
^{3}\psi =\overline{\psi }\sigma ^{3}\theta ,  \nonumber \\
(\overline{\theta }\rho ^{a}\psi )^{*} &=&\overline{\theta }\rho ^{a}\psi =-%
\overline{\psi }\rho ^{a}\theta .  \label{A8}
\end{eqnarray}
\smallskip \smallskip It is useful to define the lower indices for the
spinorial components. These are defined as follows 
\begin{equation}
\theta _{+}=\theta ^{-},\qquad \theta _{-}=-\theta ^{+}.  \label{A9}
\end{equation}
Using this notation, the scalar quantities $\overline{\theta }\psi ,$ $%
\overline{\theta }\sigma ^{3}\psi $ in components are written as 
\begin{eqnarray*}
\overline{\theta }\psi &=&\theta ^{+}\psi _{+}+\theta ^{-}\psi _{-}, \\
\overline{\theta }\sigma ^{3}\psi &=&\theta ^{+}\psi _{+}-\theta ^{-}\psi
_{-}.
\end{eqnarray*}


\smallskip

\smallskip

\smallskip

\begin{thebibliography}{99}
\bibitem{1}  \smallskip J. Wess and J. Bagger, \textit{Supersymmetry and
supergravity, }Princeton University Press, 1983 (second edition 1992); D.
Bailin, A. love, \textit{Supersymmetric gauge field theory and string theory}%
, IoP publishing, 1994. For recent reviews see Jose Figueroa-O'Farrill, 
\textit{BUSSTEPP Lectures on Supersymmetry}, hep-th/0109172; Adel Bilal, 
\textit{Introduction to Supersymmetry}, hep-th/0101055; Michael Dine, 
\textit{String Theory, Large Dimensions and Supersymmetry }hep-th/0107263

\bibitem{2}  J. Polchinski, \textit{String Theory}, Vol.1,2, Cambridge Univ.
Press, 1998\smallskip

\bibitem{3.1}  Alain Connes, Michael R. Douglas, Albert Schwarz, \textit{%
Noncommutative Geometry and Matrix Theory, }JHEP 9802 (1998) 003,
hep-th/9711162

\bibitem{3.2}  Michael R. Douglas, Chris Hull, \textit{D-branes and the
Noncommutative Torus, }JHEP 9802 (1998) 008, hep-th/9711165

\bibitem{3.3}  Nathan Seiberg, Edward Witten, \textit{String Theory and
Noncommutative Geometry, }JHEP 9909 (1999) 032, hep-th/9908142\smallskip

\bibitem{3.4}  Nathan Seiberg, \textit{A Note on Background Independence in
Noncommutative Gauge Theories, Matrix Model and Tachyon Condensation, }JHEP
0009 (2000) 003, hep-th/0008013

\bibitem{3.5}  \smallskip \smallskip Michael R. Douglas, Nikita A.Nekrasov 
\textit{Noncommutative Field Theory, }hep-th/0106048

\bibitem{4.1}  Chong-Sun Chu, Frederic Zamora, \textit{Manifest
Supersymmetry in Non-Commutative Geometry, } JHEP 0002 (2000) 022,
hep-th/9912153

\bibitem{4.2}  Jonathan Bagger, Ioannis Giannakis, \textit{Spacetime
Supersymmetry in a nontrivial NS-NS Superstring Background, }hep-th/0107260

\bibitem{5.1}  S. Ferrara, M.A. Lled\'{o}, \textit{Some Aspects of
Deformations of Supersymmetric Field Theories, }JHEP 0005 (2000) 008,
hep-th/0002084

\bibitem{5.2}  M. A. Lled\'{o}, \textit{Supersymmetric Field Theories on
Deformed Space-Time, hep-th/0011268 }

\bibitem{5.3}  \smallskip Dietmar Klemm, Silvia Penati, Laura Tamassia, 
\textit{Non(anti)commutative Superspace, }hep-th/0104190

\bibitem{5.4}  \smallskip P. Kosinski, J. Lukierski, P. Maslanka, \textit{%
Quantum Deformations of Space-Time SUSY and Noncommutative Superfield
Theory, } hep-th/0011053

\bibitem{5.5}  I. V. Tyutin, \textit{The general form of the star-product on
the Grassman algebra, }Journal-ref: Theor.Math.Phys. 127 (2001) 619-631;
Teor.Mat.Fiz. 127 (2001) 253-267, hep-th/0101046

\bibitem{5.6}  J. W. Moffat, \textit{Noncommutative and Non-Anticommutative
Quantum Field Theory, }Phys. Lett. B506 (2001) 193-199, hep-th/0011035

\bibitem{6.1}  Seiji Terashima,\textit{\ A Note on Superfields and
Noncommutative Geometry,}Phys. Lett. B482 (2000) 276-282, hep-th/0002119

\bibitem{6.2}  \smallskip H. O. Girotti, M. Gomes, V. O. Rivelles, A. J. da
Silva,\textit{A Consistent Noncommutative Field Theory: the Wess-Zumino
Model, } Nucl.Phys. B587 (2000) 299-310, hep-th/0005272

\bibitem{6.3}  \smallskip A. A. Bichl, J. M. Grimstrup, H. Grosse, L. Popp,
M. Schweda, R. Wulkenhaar, \textit{The Superfield Formalism Applied to the
Noncommutative Wess-Zumino Model, } JHEP 0010 (2000) 046, hep-th/0007050

\bibitem{6.4}  \smallskip Teparksorn Pengpan, Xiaozhen Xiong, \textit{A Note
on the Non-Commutative Wess-Zumino Model, }Phys.Rev. D63 (2001) 085012,
hep-th/0009070

\bibitem{6.5}  Daniela Zanon\textit{,\ Noncommutative perturbation in
superspace, }Phys.Lett. B504 (2001) 101-108, hep-th/0009196

\bibitem{6.6}  I. L. Buchbinder, M. Gomes, A. Yu. Petrov, V. O. Rivelles, 
\textit{Superfield Effective Action in the Noncommutative Wess-Zumino Model, 
}hep-th/0107022

\bibitem{6.11}  \smallskip Adi Armoni, Ruben Minasian, Stefan Theisen, 
\textit{On non-commutative N=2 super Yang-Mills, }Phys.Lett. B513 (2001)
406-412, hep-th/0102007

\bibitem{6.12}  \smallskip Daniela Zanon,\textit{\ Noncommutative $\mathcal{N%
}=1,2$ super U(N) Yang-Mills: UV/IR mixing and effective action results at
one loop,} Phys.Lett. B502 (2001) 265-273, hep-th/0012009

\bibitem{6.13}  \smallskip \smallskip Mario Pernici, Alberto Santambrogio,
Daniela Zanon,\textit{\ The one-loop effective action of noncommutative $%
\mathcal{N}=4$ super Yang-Mills is gauge invariant, }Phys.Lett. B504 (2001)
131-140, hep-th/0011140

\bibitem{6.10}  \smallskip N.Grandi, R.L.Pakman, F.A.Schaposnik,\textit{\
Supersymmetric Dirac-Born-Infeld theory in noncommutative space,} Nucl.Phys.
B588 (2000) 508-520, hep-th/0004104

\bibitem{6.9}  \smallskip Sergei \thinspace V. Ketov,\textit{\ N=2
super-Born-Infeld theory revisited, }Class.Quant.Grav. 17 (2000) L91,
hep-th/0005126

\bibitem{6.8}  \smallskip \smallskip H. O. Girotti, M. Gomes, V. O.
Rivelles, A. J. da Silva, \textit{The Noncommutative Supersymmetric
Nonlinear Sigma Model, }hep-th/0102101, hep-th/0011140

\bibitem{6.7}  \smallskip Victor O. Rivelles,\textit{\ Noncommutative
Supersymmetric Field Theories,} Braz.J.Phys. 31 (2001) 255-262,
hep-th/0103131

\bibitem{6.15}  Shiraz Minwalla, Mark Van Raamsdonk, Nathan Seiberg, \textit{%
Noncommutative Perturbative Dynamics, }JHEP 0002 (2000) 020, hep-th/9912072

\bibitem{10}  Rajesh Gopakumar, Shiraz Minwalla, Andrew Strominger, \textit{%
Noncommutative Solitons, }JHEP 0005 (2000) 020, hep-th/0003160

\bibitem{7.1}  G. V. Dunne, R. Jackiw, and C. A. Trugenberger, ``\textit{%
Topological'' (Chern-Simons) quantum mechanics,} Phys. Rev. D41, 1990, 661

\bibitem{7.2}  Daniela Bigatti, Leonard Susskind, \textit{Magnetic fields,
branes and noncommutative geometry, }Phys.Rev. D62 (2000) 066004,
hep-th/9908056

\bibitem{7.3}  V.P. Nair, A.P. Polychronakos, \textit{Quantum mechanics on
the noncommutative plane and sphere, }Phys.Lett. B505 (2001) 267-274,
hep-th/0011172

\bibitem{7.4}  Corneliu Sochichiu, \textit{A Note on Noncommutative and
False Noncommutative spaces, }hep-th/0010149

\bibitem{7.5}  A.Smailagic, E.Spallucci, \textit{New isotropic vs
anisotropic phase of non-commutative 2D harmonic oscillator, }hep-th/0108216

\bibitem{7.6}  Stefano Bellucci, Armen Nersessian, Corneliu Sochichiu, 
\textit{\ Two phases of the noncommutative quantum mechanics, }hep-th/0106138

\bibitem{7.7}  Agapitos Hatzinikitas, Ioannis Smyrnakis, \textit{The
noncommutative harmonic oscillator in more than one dimensions, }%
hep-th/0103074

\bibitem{7.8}  Bogdan Morariu, Alexios P. Polychronakos, \textit{Quantum
Mechanics on the Noncommutative Torus, }Nucl.Phys.B610:531-544,2001\textit{, 
}hep-th/0102157

\bibitem{9}  Lorenzo Cornalba, Ricardo Schiappa,\textit{\ Nonassociative
Star Product Deformations for D-brane Worldvolumes in Curved Backgrounds, }%
hep-th/0101219

\bibitem{8}  R. Abbaspur, S. Parvizi, \textit{Noncommutative Deformation of
the }$\mathcal{N}=1,\,D=4$\textit{\ Supertranslation Algebra }(unpublished)

\bibitem{11}  R. Abbaspur, \textit{Generalized Noncommutative Supersymmetry
from a New Gauge Symmetry,} hep-th/0206170
\end{thebibliography}
\end{document}